\documentclass[twocolumn,letterpaper,aps,prd,longbibliography,superscriptaddress,nofootinbib,floatfix]{revtex4-1}

\usepackage{graphicx}	
\usepackage{amsmath}
\usepackage{pifont}
\usepackage{multirow}
\usepackage{xcolor}
\usepackage[colorinlistoftodos]{todonotes}
\usepackage{xspace}	

\newcommand{\pt}{\mbox{$p_T$}\xspace}
\newcommand {\meanptsq}  {\ensuremath{\langle p_{\mathrm{\textsc{t}}}^{2} \rangle}\xspace}
\newcommand{\gevcc}{\mbox{GeV/$c^{2}$}\xspace}

\newcommand{\meanpt}{\mbox{$\langle p_T \rangle$}\xspace}

\newcommand{\sqrts}{\mbox{$\sqrt{s}$}\xspace}
\newcommand{\jpsi}{\mbox{$J/\psi$}\xspace}
\newcommand{\psip}{\mbox{$\psi(2S)$}\xspace}

\newcommand{\pp}{\mbox{$p$+$p$}\xspace}

\newcommand{\cc}{$c\bar{c}$\xspace}

\newcommand{\ccbar}{\ensuremath{c\bar{c}}}

\newcommand{\accEff}{$A\varepsilon_{\rm rec}$\xspace}

\begin{document}

\title{$J/\psi$ and $\psi(2S)$ production at forward rapidity in
$p$$+$$p$ collisions at $\sqrt{s}=510$ GeV}

\newcommand{\abilene}{Abilene Christian University, Abilene, Texas 79699, USA}
\newcommand{\augie}{Department of Physics, Augustana University, Sioux Falls, South Dakota 57197, USA}
\newcommand{\banaras}{Department of Physics, Banaras Hindu University, Varanasi 221005, India}
\newcommand{\barc}{Bhabha Atomic Research Centre, Bombay 400 085, India}
\newcommand{\baruch}{Baruch College, City University of New York, New York, New York, 10010 USA}
\newcommand{\bnlcoll}{Collider-Accelerator Department, Brookhaven National Laboratory, Upton, New York 11973-5000, USA}
\newcommand{\bnlphys}{Physics Department, Brookhaven National Laboratory, Upton, New York 11973-5000, USA}
\newcommand{\caucr}{University of California-Riverside, Riverside, California 92521, USA}
\newcommand{\charlesczech}{Charles University, Ovocn\'{y} trh 5, Praha 1, 116 36, Prague, Czech Republic}
\newcommand{\cns}{Center for Nuclear Study, Graduate School of Science, University of Tokyo, 7-3-1 Hongo, Bunkyo, Tokyo 113-0033, Japan}
\newcommand{\colorado}{University of Colorado, Boulder, Colorado 80309, USA}
\newcommand{\columbia}{Columbia University, New York, New York 10027 and Nevis Laboratories, Irvington, New York 10533, USA}
\newcommand{\czechtech}{Czech Technical University, Zikova 4, 166 36 Prague 6, Czech Republic}
\newcommand{\debrecen}{Debrecen University, H-4010 Debrecen, Egyetem t{\'e}r 1, Hungary}
\newcommand{\elte}{ELTE, E{\"o}tv{\"o}s Lor{\'a}nd University, H-1117 Budapest, P{\'a}zm{\'a}ny P.~s.~1/A, Hungary}
\newcommand{\eszterhazy}{Eszterh\'azy K\'aroly University, K\'aroly R\'obert Campus, H-3200 Gy\"ongy\"os, M\'atrai \'ut 36, Hungary}
\newcommand{\ewha}{Ewha Womans University, Seoul 120-750, Korea}
\newcommand{\famu}{Florida A\&M University, Tallahassee, FL 32307, USA}
\newcommand{\fsu}{Florida State University, Tallahassee, Florida 32306, USA}
\newcommand{\gsu}{Georgia State University, Atlanta, Georgia 30303, USA}
\newcommand{\hanyang}{Hanyang University, Seoul 133-792, Korea}
\newcommand{\hiroshima}{Hiroshima University, Kagamiyama, Higashi-Hiroshima 739-8526, Japan}
\newcommand{\howard}{Department of Physics and Astronomy, Howard University, Washington, DC 20059, USA}
\newcommand{\ihepprot}{IHEP Protvino, State Research Center of Russian Federation, Institute for High Energy Physics, Protvino, 142281, Russia}
\newcommand{\illuiuc}{University of Illinois at Urbana-Champaign, Urbana, Illinois 61801, USA}
\newcommand{\inrras}{Institute for Nuclear Research of the Russian Academy of Sciences, prospekt 60-letiya Oktyabrya 7a, Moscow 117312, Russia}
\newcommand{\instpasczech}{Institute of Physics, Academy of Sciences of the Czech Republic, Na Slovance 2, 182 21 Prague 8, Czech Republic}
\newcommand{\isu}{Iowa State University, Ames, Iowa 50011, USA}
\newcommand{\jaea}{Advanced Science Research Center, Japan Atomic Energy Agency, 2-4 Shirakata Shirane, Tokai-mura, Naka-gun, Ibaraki-ken 319-1195, Japan}
\newcommand{\jeonbuk}{Jeonbuk National University, Jeonju, 54896, Korea}
\newcommand{\jyvaskyla}{Helsinki Institute of Physics and University of Jyv{\"a}skyl{\"a}, P.O.Box 35, FI-40014 Jyv{\"a}skyl{\"a}, Finland}
\newcommand{\kek}{KEK, High Energy Accelerator Research Organization, Tsukuba, Ibaraki 305-0801, Japan}
\newcommand{\korea}{Korea University, Seoul, 02841}
\newcommand{\kurchatov}{National Research Center ``Kurchatov Institute", Moscow, 123098 Russia}
\newcommand{\kyoto}{Kyoto University, Kyoto 606-8502, Japan}
\newcommand{\lahorelums}{Physics Department, Lahore University of Management Sciences, Lahore 54792, Pakistan}
\newcommand{\lawllnl}{Lawrence Livermore National Laboratory, Livermore, California 94550, USA}
\newcommand{\losalamos}{Los Alamos National Laboratory, Los Alamos, New Mexico 87545, USA}
\newcommand{\lund}{Department of Physics, Lund University, Box 118, SE-221 00 Lund, Sweden}
\newcommand{\lyon}{IPNL, CNRS/IN2P3, Univ Lyon, Université Lyon 1, F-69622, Villeurbanne, France}
\newcommand{\maryland}{University of Maryland, College Park, Maryland 20742, USA}
\newcommand{\mass}{Department of Physics, University of Massachusetts, Amherst, Massachusetts 01003-9337, USA}
\newcommand{\michigan}{Department of Physics, University of Michigan, Ann Arbor, Michigan 48109-1040, USA}
\newcommand{\muhlenberg}{Muhlenberg College, Allentown, Pennsylvania 18104-5586, USA}
\newcommand{\myongji}{Myongji University, Yongin, Kyonggido 449-728, Korea}
\newcommand{\nara}{Nara Women's University, Kita-uoya Nishi-machi Nara 630-8506, Japan}
\newcommand{\natmephi}{National Research Nuclear University, MEPhI, Moscow Engineering Physics Institute, Moscow, 115409, Russia}
\newcommand{\newmex}{University of New Mexico, Albuquerque, New Mexico 87131, USA}
\newcommand{\nmsu}{New Mexico State University, Las Cruces, New Mexico 88003, USA}
\newcommand{\northcg}{Physics and Astronomy Department, University of North Carolina at Greensboro, Greensboro, North Carolina 27412, USA}
\newcommand{\ohio}{Department of Physics and Astronomy, Ohio University, Athens, Ohio 45701, USA}
\newcommand{\ornl}{Oak Ridge National Laboratory, Oak Ridge, Tennessee 37831, USA}
\newcommand{\orsay}{IPN-Orsay, Univ.~Paris-Sud, CNRS/IN2P3, Universit\'e Paris-Saclay, BP1, F-91406, Orsay, France}
\newcommand{\peking}{Peking University, Beijing 100871, People's Republic of China}
\newcommand{\pnpi}{PNPI, Petersburg Nuclear Physics Institute, Gatchina, Leningrad region, 188300, Russia}
\newcommand{\pusan}{Pusan National University, Busan, 46241, South Korea}
\newcommand{\riken}{RIKEN Nishina Center for Accelerator-Based Science, Wako, Saitama 351-0198, Japan}
\newcommand{\rikjrbrc}{RIKEN BNL Research Center, Brookhaven National Laboratory, Upton, New York 11973-5000, USA}
\newcommand{\rikkyo}{Physics Department, Rikkyo University, 3-34-1 Nishi-Ikebukuro, Toshima, Tokyo 171-8501, Japan}
\newcommand{\saispbstu}{Saint Petersburg State Polytechnic University, St.~Petersburg, 195251 Russia}
\newcommand{\seoulnat}{Department of Physics and Astronomy, Seoul National University, Seoul 151-742, Korea}
\newcommand{\stonybrkc}{Chemistry Department, Stony Brook University, SUNY, Stony Brook, New York 11794-3400, USA}
\newcommand{\stonycrkp}{Department of Physics and Astronomy, Stony Brook University, SUNY, Stony Brook, New York 11794-3800, USA}
\newcommand{\tenn}{University of Tennessee, Knoxville, Tennessee 37996, USA}
\newcommand{\titech}{Department of Physics, Tokyo Institute of Technology, Oh-okayama, Meguro, Tokyo 152-8551, Japan}
\newcommand{\tsukuba}{Tomonaga Center for the History of the Universe, University of Tsukuba, Tsukuba, Ibaraki 305, Japan}
\newcommand{\vandy}{Vanderbilt University, Nashville, Tennessee 37235, USA}
\newcommand{\weizmann}{Weizmann Institute, Rehovot 76100, Israel}
\newcommand{\wigner}{Institute for Particle and Nuclear Physics, Wigner Research Centre for Physics, Hungarian Academy of Sciences (Wigner RCP, RMKI) H-1525 Budapest 114, POBox 49, Budapest, Hungary}
\newcommand{\yonsei}{Yonsei University, IPAP, Seoul 120-749, Korea}
\newcommand{\zagreb}{Department of Physics, Faculty of Science, University of Zagreb, Bijeni\v{c}ka c.~32 HR-10002 Zagreb, Croatia}
\affiliation{\abilene}
\affiliation{\augie}
\affiliation{\banaras}
\affiliation{\barc}
\affiliation{\baruch}
\affiliation{\bnlcoll}
\affiliation{\bnlphys}
\affiliation{\caucr}
\affiliation{\charlesczech}
\affiliation{\cns}
\affiliation{\colorado}
\affiliation{\columbia}
\affiliation{\czechtech}
\affiliation{\debrecen}
\affiliation{\elte}
\affiliation{\eszterhazy}
\affiliation{\ewha}
\affiliation{\famu}
\affiliation{\fsu}
\affiliation{\gsu}
\affiliation{\hanyang}
\affiliation{\hiroshima}
\affiliation{\howard}
\affiliation{\ihepprot}
\affiliation{\illuiuc}
\affiliation{\inrras}
\affiliation{\instpasczech}
\affiliation{\isu}
\affiliation{\jaea}
\affiliation{\jeonbuk}
\affiliation{\jyvaskyla}
\affiliation{\kek}
\affiliation{\korea}
\affiliation{\kurchatov}
\affiliation{\kyoto}
\affiliation{\lahorelums}
\affiliation{\lawllnl}
\affiliation{\losalamos}
\affiliation{\lund}
\affiliation{\lyon}
\affiliation{\maryland}
\affiliation{\mass}
\affiliation{\michigan}
\affiliation{\muhlenberg}
\affiliation{\myongji}
\affiliation{\nara}
\affiliation{\natmephi}
\affiliation{\newmex}
\affiliation{\nmsu}
\affiliation{\northcg}
\affiliation{\ohio}
\affiliation{\ornl}
\affiliation{\orsay}
\affiliation{\peking}
\affiliation{\pnpi}
\affiliation{\pusan}
\affiliation{\riken}
\affiliation{\rikjrbrc}
\affiliation{\rikkyo}
\affiliation{\saispbstu}
\affiliation{\seoulnat}
\affiliation{\stonybrkc}
\affiliation{\stonycrkp}
\affiliation{\tenn}
\affiliation{\titech}
\affiliation{\tsukuba}
\affiliation{\vandy}
\affiliation{\weizmann}
\affiliation{\wigner}
\affiliation{\yonsei}
\affiliation{\zagreb}
\author{U.A.~Acharya} \affiliation{\gsu} 
\author{A.~Adare} \affiliation{\colorado} 
\author{C.~Aidala} \affiliation{\michigan} 
\author{N.N.~Ajitanand} \altaffiliation{Deceased} \affiliation{\stonybrkc} 
\author{Y.~Akiba} \email[PHENIX Spokesperson: ]{akiba@rcf.rhic.bnl.gov} \affiliation{\riken} \affiliation{\rikjrbrc} 
\author{R.~Akimoto} \affiliation{\cns} 
\author{M.~Alfred} \affiliation{\howard} 
\author{N.~Apadula} \affiliation{\isu} \affiliation{\stonycrkp} 
\author{Y.~Aramaki} \affiliation{\riken} 
\author{H.~Asano} \affiliation{\kyoto} \affiliation{\riken} 
% ...
\collaboration{PHENIX Collaboration} \noaffiliation

\date{\today}


\begin{abstract}

The PHENIX experiment at the Relativistic Heavy Ion Collider has 
measured the differential cross section, mean transverse momentum, mean 
transverse momentum squared of inclusive $J/\psi$ and cross-section 
ratio of $\psi(2S)$ to $J/\psi$ at forward rapidity in \pp collisions at 
\sqrts = 510 GeV via the dimuon decay channel. Comparison is made to 
inclusive $J/\psi$ cross sections measured at \sqrts = 200 GeV and 
2.76--13 TeV.  The result is also compared to leading-order 
nonrelativistic QCD calculations coupled to a color-glass-condensate 
description of the low-$x$ gluons in the proton at low transverse 
momentum ($p_T$) and to next-to-leading order nonrelativistic QCD 
calculations for the rest of the $p_T$ range. These calculations 
overestimate the data at low $p_T$.  While consistent with the data 
within uncertainties above $\approx3$ GeV/$c$, the calculations are 
systematically below the data. The total cross section times the 
branching ratio is BR $d\sigma^{J/\psi}_{pp}/dy (1.2<|y|<2.2,
0<\pt<10~\mbox{GeV/$c$}) =$ 54.3 $\pm$ 0.5 (stat) $\pm$ 5.5 (syst) nb.

\end{abstract}

\maketitle

\section{Introduction}
\label{Sec:intro}

Charmonium states such as \jpsi and \psip mesons are bound states of a 
charm and anti-charm quark ($\ccbar$). At the Relativistic Heavy Ion 
Collider (RHIC) energies, they are produced mostly from hard scattering 
of two gluons into a $\ccbar$ pair followed by the evolution of this 
pair through a hadronization process to form a physical charmonium. 
Despite several decades of extensive 
studies~\cite{PhysRevLett.69.3704,PhysRevLett.98.232002,PhysLett.B.704.442,PhysLett.B.718.295,EurPhysJC.76.184,EuroPhysJC.77.392,PhysRevD.100.052009,PhysRevLett.106.042002,PhysRevLett.113.192301} 
since the discovery of \jpsi, we still have very limited knowledge about 
the \jpsi production mechanism and hadronization. Therefore, carrying out as 
many charmonium measurements as possible in \pp collisions over a wide 
range of transverse momentum (\pt) and of rapidity ($y$) at different 
energies is essential to understanding production mechanisms.  These 
measurements over a wide range of \pt (down to zero \pt) and rapidity 
allow calculating quantities, such as the mean transverse momentum 
\meanpt, the mean transverse momentum squared \meanptsq, and the 
\pt-integrated cross section $d\sigma/dy$. The collision energy 
dependence of these quantities can put stringent constraints on the 
different theoretical approaches that are used to describe the hadronic 
production of \jpsi. These approaches include the color-evaporation 
model (CEM)~\cite{PhysLettB.67.217,PhysLettB.390.323}, the color-singlet 
model (CSM)~\cite{PhysLettB.102.364} and the nonrelativistic quantum 
chromodynamics formalism (NRQCD)~\cite{PhysRevD.51.1125}. In this work, 
we compare the data to NRQCD, an effective field theory derived from QCD 
and valid for heavy-quark pairs with low relative velocity, where a 
\jpsi can be formed from \cc pair produced in a color-singlet or a 
color-octet state.

In this paper, we present the inclusive \jpsi production cross section 
and the ratio of \psip to \jpsi production cross sections at forward 
rapidity ($1.2<|y|<2.2$) measured in \pp collisions at center of mass 
energy \sqrts = 510~GeV. These mesons are measured in the dimuon decay 
channel. The \jpsi inclusive differential cross sections are obtained as 
a function of \pt and $y$ over a wide range of \pt. The \jpsi and \psip 
results at \sqrts = 510~GeV are the first measurements at this rapidity. 
Comparisons to similar PHENIX measurements performed at \sqrts = 200 
GeV~\cite{PhysRevLett.98.232002} and Large Hadron Collider (LHC) 
measurements at \sqrts = 2.76, 5.02, 7, 8 and 13 
TeV~\cite{PhysLett.B.704.442,PhysLett.B.718.295,EurPhysJC.76.184,EuroPhysJC.77.392} 
allow studying the variations of \meanpt, \meanptsq and $d\sigma/dy$ as 
a function of \sqrts. The results are also compared to next-to-leading 
order (NLO) NRQCD calculations~\cite{PhysRevLett.106.042002}.

The paper is organized as follows: the PHENIX apparatus is described in 
Sec.~\ref{sec:detector}, the data samples used for this analysis and the 
analysis procedure are discussed in Sec.~\ref{sec:analysis}, while the 
results are presented and compared to measurements at different \sqrts 
as well as to models in Sec.~\ref{sec:results}.

\section{Experimental Setup}
\label{sec:detector}

A complete description of the PHENIX detector can be found in 
Ref.~\cite{NIMA.499.469}.  Only the detector systems relevant to this 
measurement are briefly described here.

\begin{figure}[htp!]
\centering
\includegraphics[angle=0, width=0.99\linewidth,trim={0 10 0 413},clip]{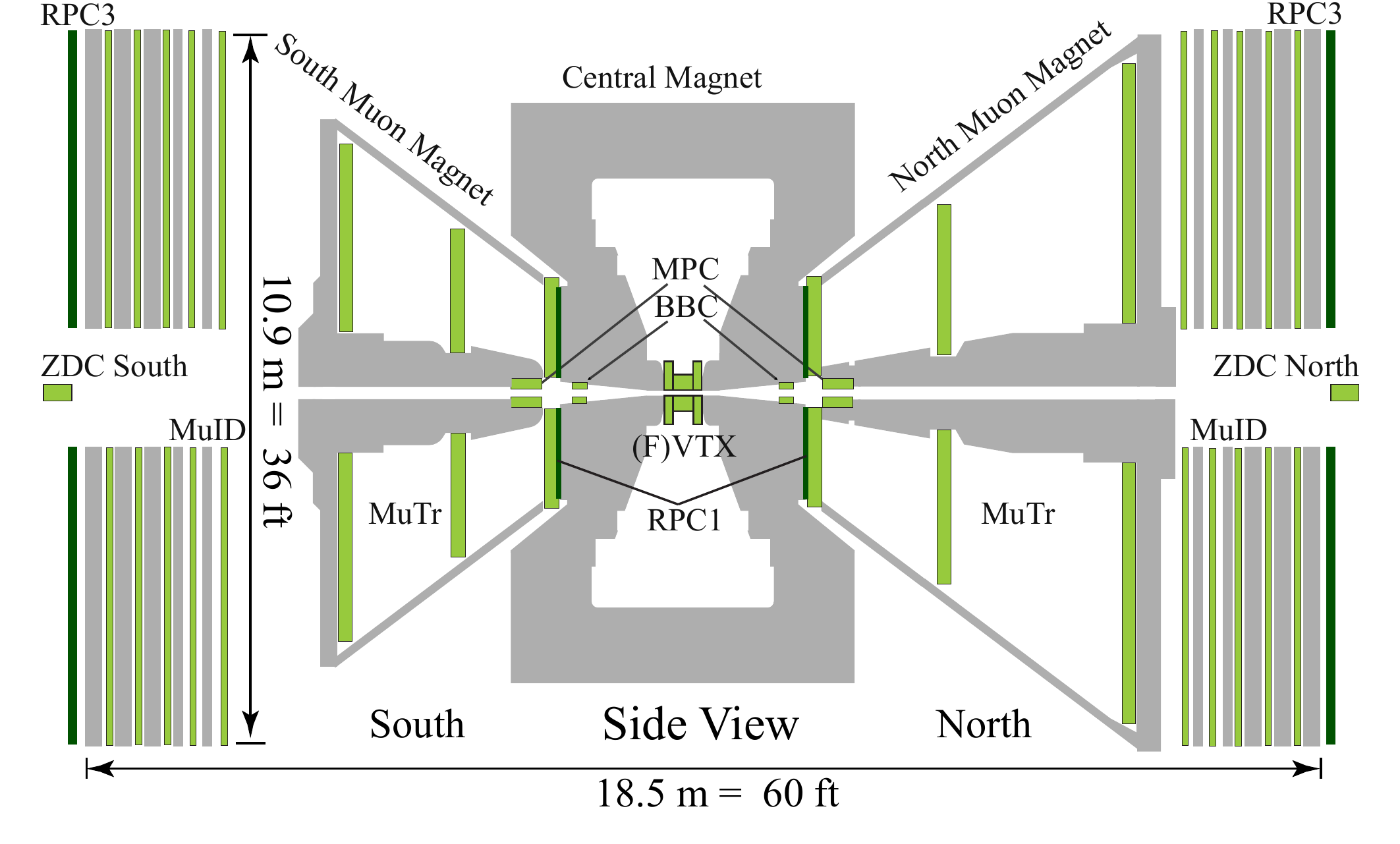}
\caption{\label{fig:Detector} A side view of the PHENIX detector, 
concentrating on the muon arm instrumentation.}
\end{figure}

The PHENIX muon spectrometers, see Fig.~\ref{fig:Detector}, cover the full 
aziumth and the north (south) arm cover forward (backward) rapidity, 
$1.2<y<2.2$ $(-2.2<y<-1.2)$. Each muon spectrometer comprises a 
hadronic absorber, a magnet, a muon tracker (MuTr), and a muon identifier 
(MuID). The absorbers comprise layers of copper, iron and stainless steel 
and have about 7.2 interaction lengths. Following the absorber in each muon 
arm is the MuTr, which comprises three stations of cathode strip chambers in 
a radial magnetic field with an integrated bending power of 0.8~T$\cdot$m. 
The MuID comprises five alternating steel absorbers and Iarocci tubes. The 
composite momentum resolution, $\delta p/p$, of particles in the analyzed 
momentum range is about 5\%, independent of momentum and dominated by 
multiple scattering. Muon candidates are identified by reconstructed tracks 
in the MuTr matched to MuID tracks that penetrate through to the last MuID 
plane.

Since 2012 the PHENIX detector had a new forward vertex detector 
(FVTX)~\cite{NIMA.755.44}, which comprises four planes of silicon strip 
detectors, finely segmented in radius and coarsely segmented in azimuth. 
For the subset of muon candidate tracks passing several of these 
detector planes, this additional information was used to improve mass 
resolution by a factor of 1.5 for studying $\psi(2S)$.

Another detector system relevant to this analysis is the beam-beam 
counter (BBC), comprising two arrays of 64~\v{C}erenkov counters, located 
on both sides of the interaction point and covering the pseudorapidity 
range $3.1<|\eta|<3.9$. The BBC system was used to measure the \pp 
collision vertex position along the beam axis ($z_{\rm vtx}$), with 2 cm 
resolution, and initial collision time. It was also used to measure the 
beam luminosity and form a minimum bias (MB) trigger.


\section{Data analysis}
\label{sec:analysis}

The results presented here are based on the data sample collected by 
PHENIX during the 2013 \pp run at \sqrts = 510~GeV. The BBC counters 
provided the MB trigger, which required at least one hit 
in each of the BBCs.  Events, in coincidence with the MB trigger, 
containing a muon pair within the acceptance of the spectrometer are 
selected by the level-1 dimuon trigger (MuIDLL1-2D) requiring that at 
least two tracks penetrate through the MuID to its last layer. The data 
sample, used in this analysis, corresponds to $3.02\times10^{12}$ MB 
events or to an integrated luminosity of 94.4~pb$^{-1}$.

\subsection{Raw yield extraction}
\label{sec:RawYieldExtraction}

A set of quality cuts is applied to the data to select good \pp events 
and good muon candidates as well as to improve the signal-to-background 
ratio. Good \pp events are selected by requiring that the collision 
occurs in the fiducial interaction region $|z_{\rm vtx}| < 30$~cm as 
measured by the BBC. Each reconstructed muon candidate comprises a 
combination of reconstructed muon tracks in the MuTr and in the MuID. 
The MuTr track is required to have more than 9 hits out of the maximum 
possible of 16 while the MuID track is required to have more than 6 hits 
out of the maximum possible of 10. In addition, a cut on individual MuTr 
track $\chi^2$ of 23 is applied. The MuTr track $\chi^2$ is calculated 
from the difference between the measured hit positions of the track and 
the subsequent fit for each MuTr track. The MuTr tracks are then matched 
to the MuID tracks at the first MuID layer by applying cuts on maximum 
position and angle differences. Furthermore, there is a minimum allowed 
single muon momentum along the beam axis, $p_z$, which is reconstructed 
and energy-loss corrected at the collision vertex, of 3.0~GeV/$c$ 
corresponding to the momentum cut effectively imposed by the absorbers. 
Finally, a cut on the $\chi^2$ of the fit of the two muon tracks to the 
common vertex of the two candidate tracks near the interaction point was 
applied.

The invariant mass distribution is formed by combining muon candidate 
tracks of opposite charges (unlike-sign). In addition to the charmonium 
signal, the resulting unlike-sign dimuon spectrum includes correlated 
and uncorrelated pairs. In the \jpsi and $\psi(2S)$ region, the 
correlated pairs arise from correlated semi-muonic decays of charmed 
hadrons, beauty and the Drell-Yan process as well as light hadron 
decays. The uncorrelated pairs are mainly coming from the decays of 
light hadrons ($\pi^\pm$, $K^\pm$ and $K^0$) which decay before or after 
passing through the absorber, and form the combinatorial background.
 
The combinatorial background is estimated using two methods: The first 
one derives the combinatorial background from the mass distribution of 
the same sign (like-sign) pairs of muon candidates within the same 
event. The second method derives the combinatorial background from the 
mass distribution of the unlike-sign pairs of muon candidates from 
different events (mixed-event) of z-vertex position 
within 2 cm. The normalization of the mass distribution of the 
combinatorial background from the like-sign dimuon distributions 
($N_{++}$ and $N_{--}$) is calculated as: $N_{\rm CB} = 
2\sqrt{N_{++}N_{--}}$. The mixed-event like-sign dimuon mass 
distribution is normalized to the same-event like-sign combinatorial 
background distribution in the invariant mass range $2.0 - 4.5$~\gevcc. 
This factor is then used to normalize the mixed-event unlike-sign dimuon 
mass distribution.

\begin{figure*}[ht!]
\includegraphics[width=0.99\linewidth]{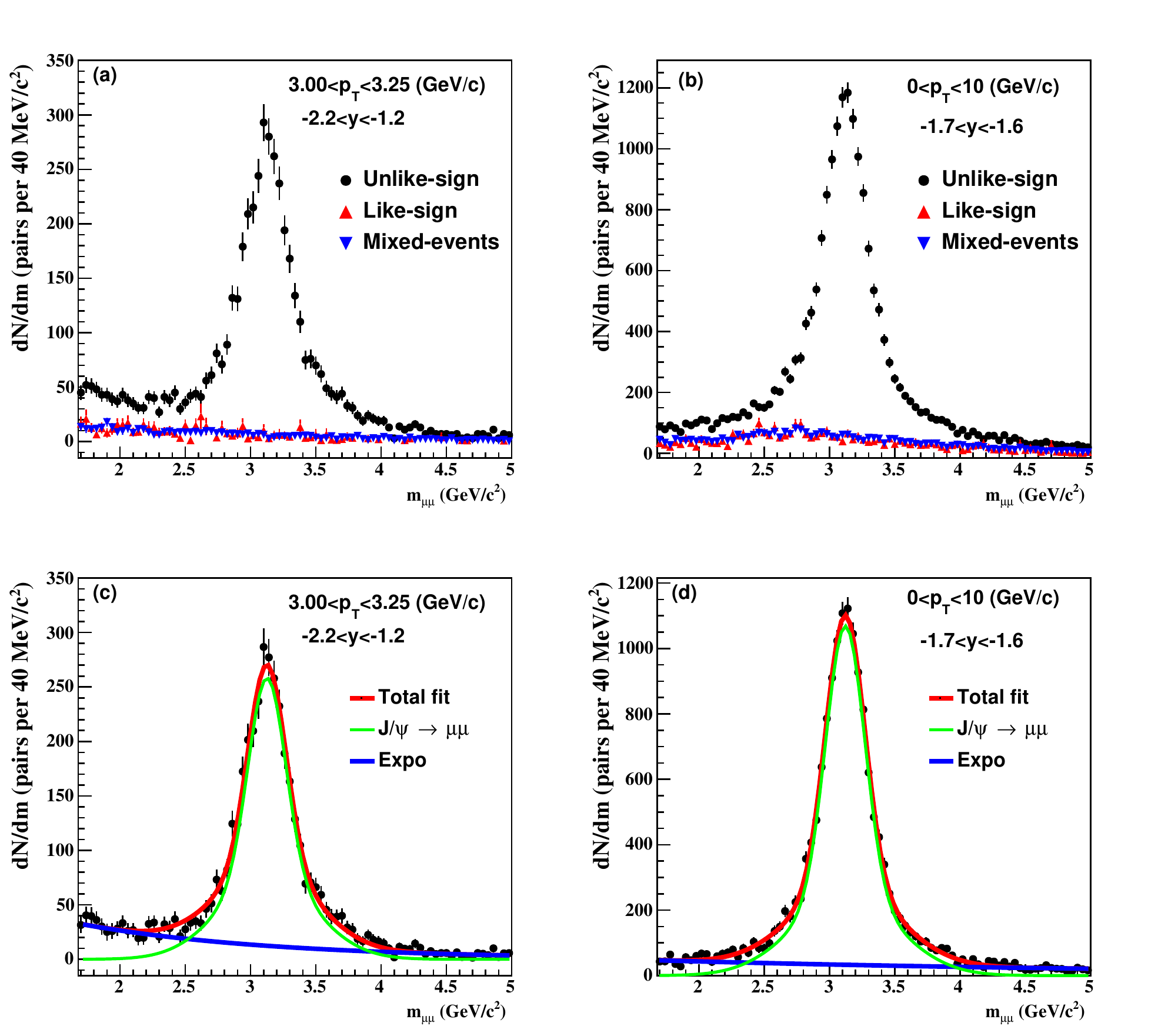}%
\caption{Raw unlike-sign dimuon spectra (closed [black] circles) along with 
normalized like-sign background (upward [red] triangles) and normalized 
mixed-event background (inverted [blue] triangles) for (a) one $p_T$ bin 
and (b) one rapidity bin. Panels (c) and (d) show the 
background-subtracted spectra fitted as described in the text for 
(c) one $p_T$ bin and (d) one rapidity bin.\label{fig:rawMassSpectra}}
\end{figure*}

Figure~\ref{fig:rawMassSpectra} shows the unlike-sign dimuon spectrum 
together with the combinatorial background estimated by both methods. 
Both background distributions from the mixed-event and like-sign methods 
are consistent, however, the mixed-event background is more 
statistically stable, because we mix each event with the 
previous four events. Therefore, the mixed-event background was used to 
subtract the uncorrelated background from the unlike-sign dimuon 
spectrum.
 
After subtracting the uncorrelated background, the unlike-sign spectra 
including the correlated background are fitted by the following 
function,

\begin{equation}
\label{eqn:fitfunc}
\begin{split}
  f(m_{\mu\mu}) = p_0 [\frac{(1-p_3)}{\sqrt{2 \pi}p_2} \mbox{exp}(-\frac{1}{2}\frac{(m_{\mu\mu}-p_1)^2}{p^2_2}) + \\
\frac{p_3}{\sqrt{2 \pi}p_4} \mbox{exp}(-\frac{1}{2}\frac{(m_{\mu\mu}-p_1)^2}{p^2_4}) ] + \\ p_5\mbox{exp}(p_6+p_7m_{\mu\mu}),
\end{split}
\end{equation}
where $p_0 - p_7$ are free parameters and $m_{\mu\mu}$ is the 
unlike-sign dimuon mass. The $J/\psi$ shape is better described with two 
Gaussian distributions, corresponding to the first two terms in 
Eq.~\ref{eqn:fitfunc}, one for the $J/\psi$ peak and a second one with 
larger width to account for the wider tails, which occurs due to 
limitations in MuTr resolution, as discussed in sec.~\ref{sec:detector}.  
The peak also includes contribution from \psip, which is not resolved. 
An exponential is used to describe the continuum contributions from 
correlated backgrounds. Panels (a) and (b) of 
Fig.~\ref{fig:rawMassSpectra} show the raw spectra for selected \pt and 
rapidity bins and panels (c) and (d) show the spectra after subtracting 
the combinatorial background fitted with the function described above 
for those selected bins.

\begin{figure}[ht!]
\includegraphics[width=1.0\linewidth]{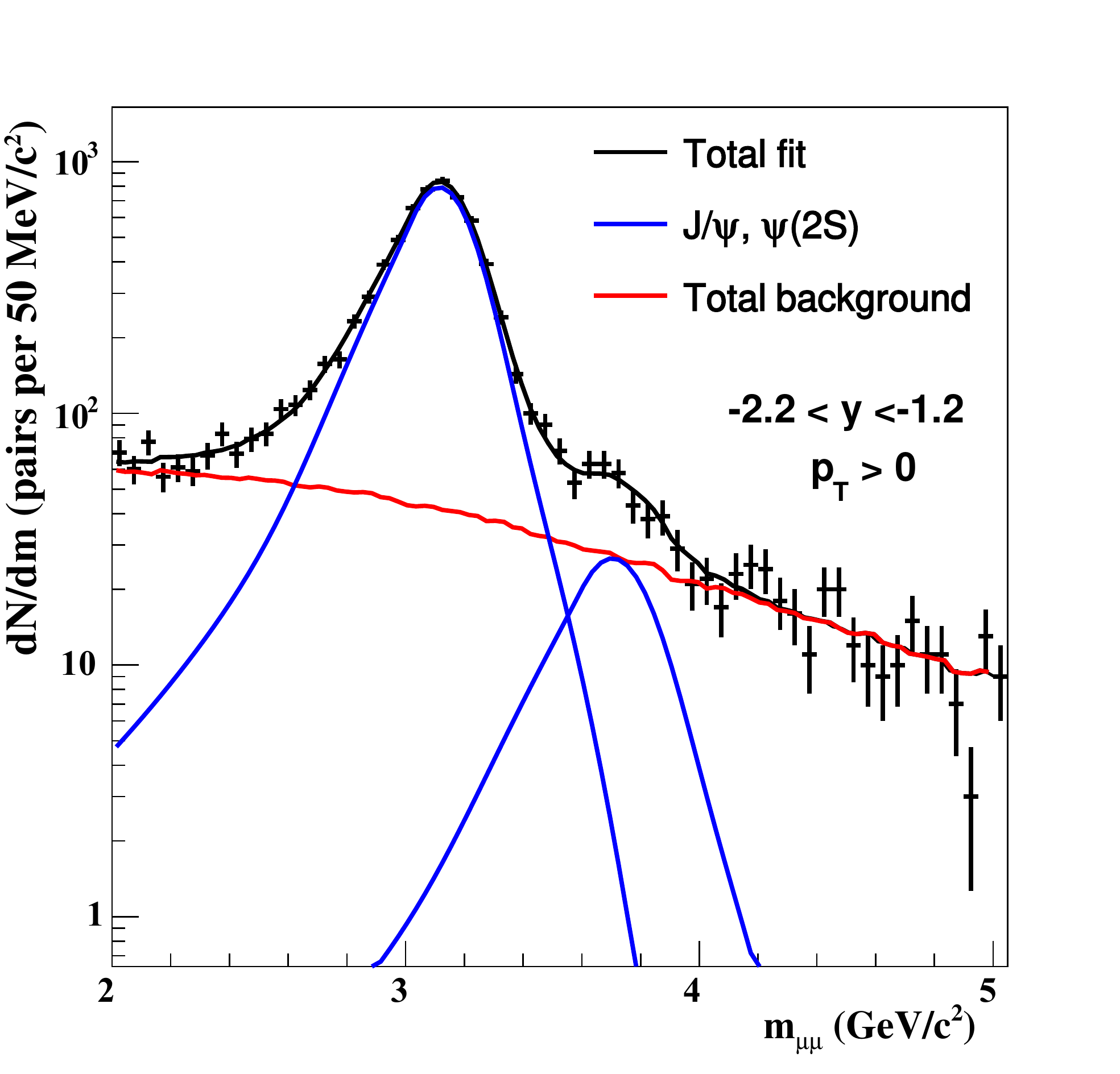}%
\caption{Raw unlike-sign dimuon spectrum summed over \pt and the whole 
backward rapidity range, $-2.2<y<-1.2$.\label{fig:rawMassSpectra_psi}}
 \end{figure}

To extract the \psip signal we improve the mass resolution of the muon 
tracking systems by utilizing the FVTX. The FVTX being located before 
the absorber allows measuring the dimuon opening angle before any 
multiple scattering occurs in the absorber~\cite{NIMA.755.44}. Using 
this additional tracking information gives a more precise measurement of 
the dimuon opening angle and thereby a more precise measurement of the 
pair mass, as well as rejecting backgrounds from decay muons that emerge 
from the absorber. However, these additional requirements on the dimuon 
tracks that are necessary to separate the $J/\psi$ and $\psi(2S)$ peaks 
also reduce the statistics by a factor of 6 due to the geometric 
acceptance of FVTX, therefore, we study the dimuon mass spectra in each 
arm integrated over \pt and rapidity within each arm. The dimuon mass 
spectrum extracted including the FVTX after subtracting the mixed-event 
background is shown in Fig.~\ref{fig:rawMassSpectra_psi}.

Given the resolution enhancement, the sum of a Gaussian and a 
crystal-ball function~\cite{crystalball,PhysRevC.95.034904}, rather than a 
double Gaussian, was used for each of \jpsi and \psip peaks to fit the 
dimuon mass spectrum. The \psip peak is expected to be wider than the 
\jpsi peak, due to the fact that the higher mass and harder \pt spectrum 
of the \psip state will produce higher momentum decay muons which have 
larger uncertainty in their reconstructed momentum in the spectrometer 
due to a smaller bend in the magnetic field. By selecting only poorly 
reconstructed tracks, we found a \jpsi width of $\approx200$ MeV/c$^2$, 
therefore, the width of the second Gaussian in the fit to the entire 
sample of tracks is set to 200 MeV/c$^2$. The ratio of widths of the 
\psip to \jpsi is set to 1.15, following expectations of the performance 
of the muon tracking system~\cite{NIMA.499.537}. The difference between 
the centroids of the \psip and \jpsi peaks is set to the Particle Data 
Group value of 589 MeV/c$^2$~\cite{ChinPhysC.38.090001}. The relative 
normalization of the second Gaussian is fixed to be the same for both 
resonances, as are the parameters for the crystal-ball line shape.

\subsection{Detector acceptance and reconstruction efficiency}

The acceptance and reconstruction efficiency 
($A\varepsilon_\mathrm{rec}$) of the muon spectrometers, including the 
MuID trigger efficiency, is determined by running a 
{\sc pythia}\footnote{We used {\sc pythia}6.421, with 
parton distribution functions 
given by CTEQ6LL. The following parameters were modified: MSEL = 
0, MSUB(86) = 1, PARP(91) = 2.1, MSTP(51) = 10041, MDME(858,1) = 0, 
MDME(859,1) = 1, MDME(860,1) = 0, and Tune 
A.}~\cite{Phys.Commun.135.238} generated \jpsi signal through a {\sc 
geant4}-based full detector simulation~\cite{NIMA.506.250} of PHENIX. 
The simulation tuned the detector response to a set of characteristics 
(dead and hot channel maps, gains, noise, etc.) that described the 
performance of each detector subsystem. The simulated vertex 
distribution was tuned to match that of the 2013 data. The simulated 
events are reconstructed in the same manner as the data and the same 
cuts are applied as in the real data analysis.

\begin{figure}[ht!]
\includegraphics[width=1.0\linewidth]{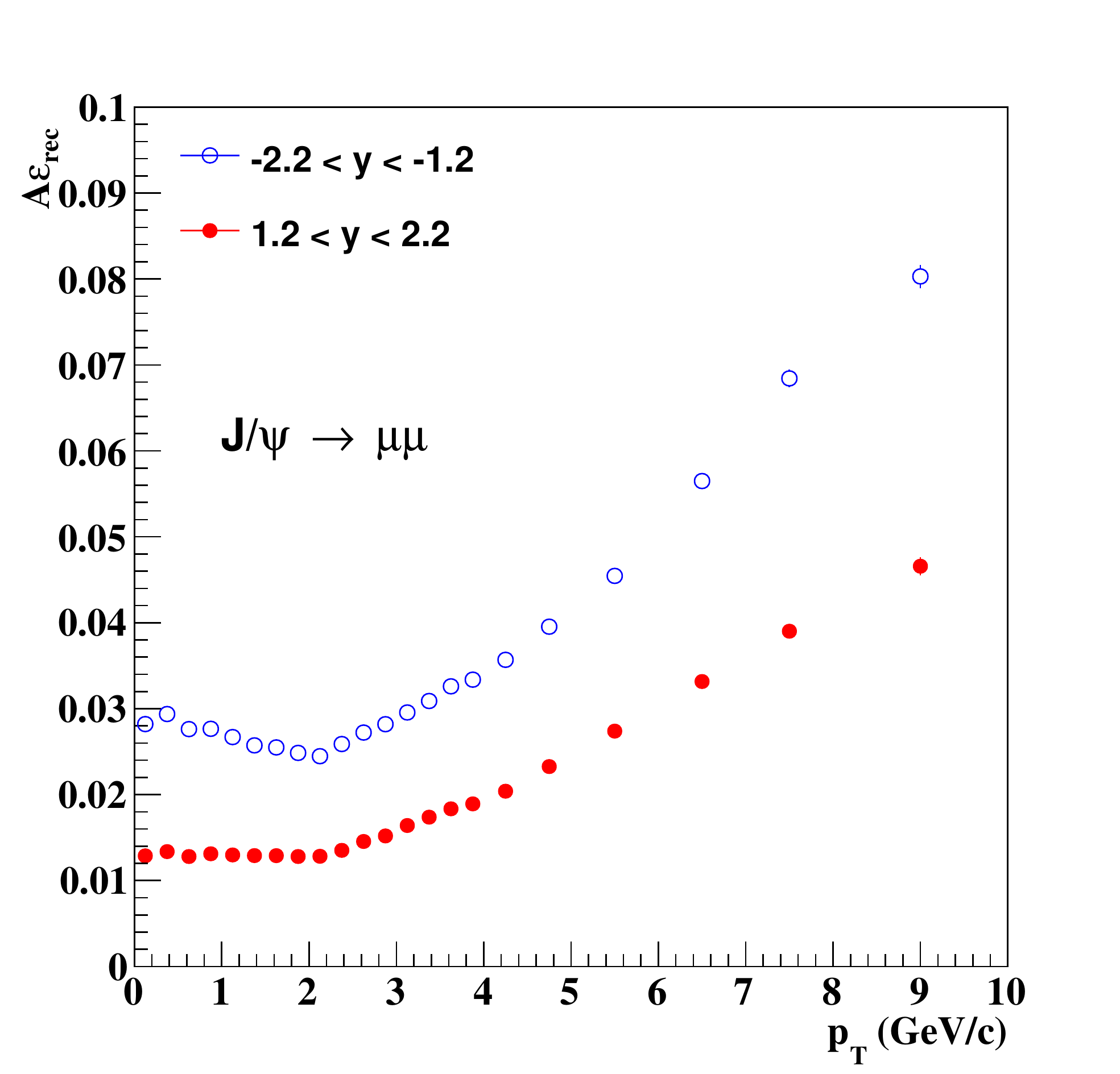}
\caption{\label{fig:AccEff} \accEff as a function of \pt for $1.2<y<2.2$ 
(closed [red] circles) and $-2.2<y<-1.2$ (open [blue] circles).}
\end{figure}

Figure~\ref{fig:AccEff} shows \accEff as a function of \pt and rapidity 
for \jpsi. The relative difference in \accEff between the two 
spectrometers is due to different detection efficiencies of the MuTr and 
MuID systems and different amount of absorber material.

In the case of \psip, we are interested in the ratio of its differential 
cross section to that of \jpsi, therefore, we extract the ratio of 
\accEff for \psip and \jpsi with addition of the FVTX information in 
analyzing the simulation to match that of the data analysis. A factor of 
0.77 (0.69) is applied to the \psip/\jpsi ratio extracted from the fit 
to the invariant mass spectrum to account for differences in acceptance, 
efficiency, and dimuon trigger efficiencies between the 
north (south) arm of the muon spectrometer.


\subsection{Differential cross section}

The differential cross section is evaluated according to the following 
relation:
\begin{equation}
\frac{d^2\sigma_\psi}{dydp_T} = \frac{1}{\Delta y\Delta p_T}\frac{N_{\psi}}{A\varepsilon_{\rm rec} BR}\frac{\sigma_{\rm BBC}}{\varepsilon_{\rm BBC}N^{\rm BBC}_{\rm MB}}
\end{equation}
where $N_{\psi}$ is the extracted \jpsi or \psip yield in $y$ and \pt 
bins with $\Delta y$ and $\Delta$\pt widths, respectively. $BR$ is the 
branching ratio where $BR_{J/\psi \rightarrow \mu^+\mu^-} = (5.93 \pm 
0.06)\times 10^{-2}$ and $BR_{\psip \rightarrow \mu^+\mu^-} = ( 7.9 \pm 
0.9 )\times 10^{-3}$~\cite{ChinPhysC.38.090001}. \accEff is the product 
of the acceptance and reconstruction efficiency. $N^{\rm BBC}_{\rm MB}$ 
$=3.02\times 10^{12}$ is the number of MB events and $\varepsilon_{\rm 
BBC}$ $= 0.91\pm 0.04$ is the efficiency of the MB trigger for events 
containing a hard scattering~\cite{PhysRevD.93.011501}. $\sigma_{\rm 
BBC}$ is the PHENIX BBC cross section, $32.5 \pm 3.2$ mb at $\sqrt{s} = 
510$ GeV, which is determined from the van der Meer scan 
technique~\cite{PhysRevLett.106.062001}.

\subsection{Systematic uncertainties}

All systematic uncertainties are evaluated as standard deviations and
are summarized in Tables~\ref{tab:sysUncer}~and~\ref{tab:sysUncerRatio}.
They are divided into three categories based upon the effect each source
has on the measured results:

\begin{description}

\item[Type-A]
Point-to-point uncorrelated uncertainties allow the data points to
move independently with respect to one another and are added in
quadrature with statistical uncertainties; however, no systematic
uncertainties of this type are associated with this measurement.

\item[Type-B] 
Point-to-point correlated uncertainties which allow the data points to move 
coherently within the quoted range to some degree. These systematic 
uncertainties include a 4\% uncertainty from MuID tube efficiency and an 
8.2\% (2.8\%) from MuTr overall efficiency for the north (south) arm. A 
3.9\% signal extraction uncertainty is assigned to account for the yield 
variations when using different functions, i.e., second, third and fourth 
order polynomials, to fit the correlated background and $\approx3\%$ 
uncertainty is assigned to account for the \psip contribution. The 
systematic uncertainty associated with \accEff includes the uncertainty on 
the input \pt and rapidity distributions which is extracted by varying these 
distributions over the range of the statistical uncertainty of the data, 
yielding 4.4\% (5.0\%) for the north (south) arm. Additional 11.2\% (8.8\%) 
systematic effect for the north (south) arm was also considered to account 
for the azimuthal angle distribution difference between data and simulation. 
To be consistent with the real data analysis, a trigger emulator was used to 
match the level-1 dimuon trigger for the data. The efficiency of the trigger 
emulator was studied by applying it to the data and comparing the resulting 
mass spectrum to the mass spectrum when applying the level-1 dimuon trigger 
which resulted in a 1.5\% (2\%) uncertainty for the north (south) arm. 
Type-B systematic uncertainties are added in quadrature and amount to 16.0\% 
(12.4\%) for the north (south) arm. They are shown as shaded bands on the 
associated data points.

\item[Type-C] 
An overall normalization uncertainty of 10\% was assigned for 
the BBC cross section and efficiency 
uncertainties~\cite{PhysRevLett.91.241803} that allow the data points 
to move together by a common multiplicative factor.

\end{description}

\begin{table}[htb!]
\caption{\label{tab:sysUncer} Systematic uncertainties associated with 
\jpsi differential cross section calculation in the north (south) arm.}
\begin{ruledtabular} \begin{tabular}{ccc}
Type- & Origin & north (south) \\
\hline
B & MuID hit efficiency & 4.0\% (4.0\%)\\
B & MuTr hit efficiency & 8.2\% (2.8\%)\\
B & Signal extraction & 3.9\% (3.9\%)\\
B & \psip contribution & 3.0\% (3.0\%)\\
B & \accEff \pt and $y$ input distributions & 4.4\% (5.0\%)\\
B & \accEff $\phi$ distribution & 11.2\% (8.8\%)\\
B & \accEff trigger emulator & 1.5\% (2.0\%)\\
\\
B & Quadratic sum & 16.0\% (12.4)\%\\
\\
C & MB trigger efficiency & 10\%\\
\end{tabular} \end{ruledtabular}
\end{table}

In the measurement of the \psip to \jpsi ratio, most of the mentioned 
systematic uncertainties cancel out.  However, the fit that was used to 
extract the yields is more complex and additional systematic 
uncertainties arose from the constraints applied during the fitting 
process.

A systematic uncertainty from constraining the normalization factor is 
determined by varying the mass range over which the factor is calculated 
and a 3\% systematic uncertainty is assigned for both arms. Systematic 
uncertainty of 3\% (7\%) was assigned to the north (south) arm on the 
fit range by varying the range around the nominal values, 2--5 GeV/$c^2$. 
The effect of constraining the second Gaussian peak width to 200 
MeV/$c^2$ was studied by varying the width between 175 and 225 
MeV/$c^2$, resulting in a systematic uncertainty of 12\% (10\%) in the 
north (south) arm.

The systematic uncertainty component on \accEff that survived the ratio 
amounts to 2.7\% (4.1\%) in the north (south) arm. The systematic 
uncertainties associated with the ratio measurement are summarized in 
Table~\ref{tab:sysUncerRatio}.
 
\begin{table}[ht!]
\caption{\label{tab:sysUncerRatio} 
Systematic uncertainties associated with the differential cross section 
ratio of \psip to \jpsi in the north (south) arm.}
\begin{ruledtabular} \begin{tabular}{ccc}
Type & Origin & north (south) \\
\hline
B & $\sigma(2s)/\sigma(1s)$ constraint & 3\% (3\%)\\
B & Background fit mass range & 3\% (7\%)\\
B & Second Gaussian width constraint & 12\% (10\%)\\
B & \accEff & 2.7\% (4.1\%)\\
\\
B & Quadratic sum & 13\% (13\%)\\
\end{tabular} \end{ruledtabular}
\end{table}

\section{Results}
\label{sec:results}

\begin{figure}[ht!]
 \includegraphics[width=1.0\linewidth]{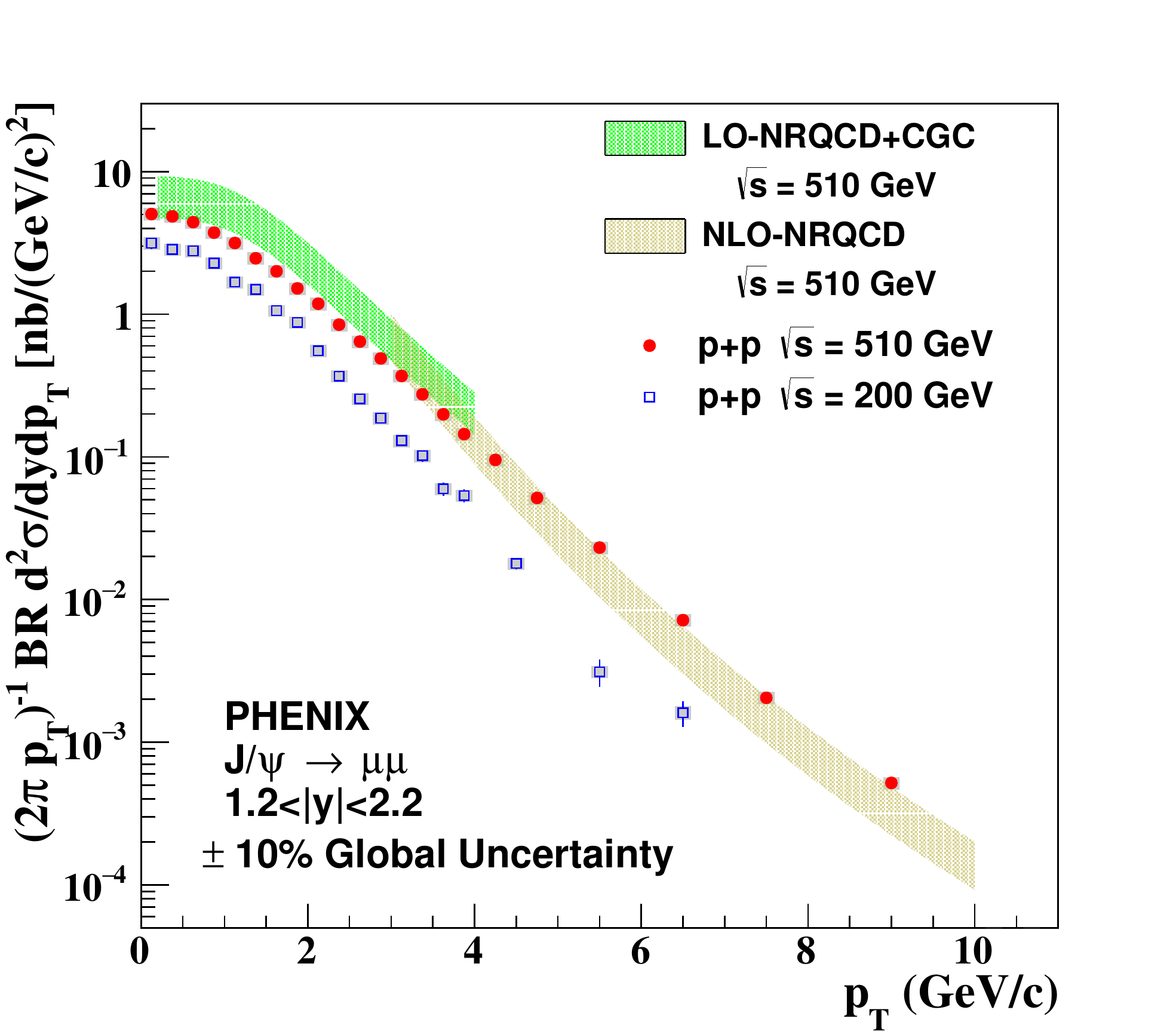}%
\caption{The inclusive \jpsi differential cross section as a function of 
\pt at $1.2<|y|<2.2$ at 510 GeV (closed [red] circles) and 
at 200 GeV (open [blue] squares). The error bars represent the 
statistical uncertainties, and gray shaded bands (although too small to 
appear on the data points) are added representing the quadratic sum of 
type-B systematic uncertainties. NRQCD calculations 
at 510 GeV~\cite{PhysRevLett.106.042002} are also shown.}
\label{fig:ptXSjpsi}
\end{figure}

The inclusive \jpsi differential cross section as a function of \pt is 
calculated independently for each muon arm, then the results are 
combined using the best-linear-unbiased-estimate 
method~\cite{EPJC.74.3004}. Results obtained using the two muon 
spectrometers are consistent within statistical 
uncertainties. The combined inclusive \jpsi differential cross section 
is shown in Fig.~\ref{fig:ptXSjpsi} and listed in 
Table~\ref{tab:ptXSjpsi}. The gray shaded bands represent the weighted 
average of the quadratic sum of type-B systematic uncertainties of the 
north and south arms, $\approx10.1\%$. The average is weighted based on the 
statistical uncertainties of each arm.

\begin{table}[htb!]
\caption{Differential cross sections in nb/(GeV/$c$)$^2$ and \pt in 
(GeV/$c$) of inclusive \jpsi with statistical and type-B systematic 
uncertainties.}
\begin{ruledtabular}\begin{tabular}{ccccc} 
& $p^{\rm min}_T$ \ 
& $p^{\rm max}_T$ \ 
& \ $\frac{BR}{2\pi p_{T}} \frac{d^{2}\sigma}{dydp_{T}}$ 
& \\
\hline
& 0.00  & \ 0.25 \ & \ $(5.04 \pm 0.23 \pm 0.51)\times10^{ 0}$ & \\
& 0.25  & \ 0.50 \ & \ $(4.85 \pm 0.17 \pm 0.49)\times10^{ 0}$ & \\
& 0.50  & \ 0.75 \ & \ $(4.42 \pm 0.15 \pm 0.45)\times10^{ 0}$ & \\
& 0.75  & \ 1.00 \ & \ $(3.73 \pm 0.13 \pm 0.38)\times10^{ 0}$ & \\
& 1.00  & \ 1.25 \ & \ $(3.16 \pm 0.11 \pm 0.32)\times10^{ 0}$ & \\
& 1.25  & \ 1.50 \ & \ $(2.47 \pm 0.08 \pm 0.25)\times10^{ 0}$ & \\
& 1.50  & \ 1.75 \ & \ $(2.00 \pm 0.07 \pm 0.20)\times10^{ 0}$ & \\
& 1.75  & \ 2.00 \ & \ $(1.52 \pm 0.05 \pm 0.15)\times10^{ 0}$ & \\
& 2.00  & \ 2.25 \ & \ $(1.18 \pm 0.04 \pm 0.12)\times10^{ 0}$ & \\
& 2.25  & \ 2.50 \ & \ $(8.45 \pm 0.30 \pm 0.85)\times10^{-1}$ & \\
& 2.50  & \ 2.75 \ & \ $(6.44 \pm 0.23 \pm 0.65)\times10^{-1}$ & \\
& 2.75  & \ 3.00 \ & \ $(4.90 \pm 0.18 \pm 0.50)\times10^{-1}$ & \\
& 3.00  & \ 3.25 \ & \ $(3.69 \pm 0.14 \pm 0.37)\times10^{-1}$ & \\
& 3.25  & \ 3.50 \ & \ $(2.74 \pm 0.10 \pm 0.28)\times10^{-1}$ & \\
& 3.50  & \ 3.75 \ & \ $(1.99 \pm 0.08 \pm 0.20)\times10^{-1}$ & \\
& 3.75  & \ 4.00 \ & \ $(1.44 \pm 0.06 \pm 0.15)\times10^{-1}$ & \\
& 4.00  & \ 4.50 \ & \ $(9.53 \pm 0.36 \pm 0.96)\times10^{-2}$ & \\
& 4.50  & \ 5.00 \ & \ $(5.16 \pm 0.21 \pm 0.52)\times10^{-2}$ & \\
& 5.00  & \ 6.00 \ & \ $(2.31 \pm 0.09 \pm 0.23)\times10^{-2}$ & \\
& 6.00  & \ 7.00 \ & \ $(7.17 \pm 0.34 \pm 0.72)\times10^{-3}$ & \\
& 7.00  & \ 8.00 \ & \ $(2.05 \pm 0.15 \pm 0.21)\times10^{-3}$ & \\
& 8.00  & \ 10.00 \ & \ $(5.18 \pm 0.44 \pm 0.52)\times10^{-4}$ & 
\end{tabular}\end{ruledtabular}
\label{tab:ptXSjpsi}
\end{table}

The data points are corrected to account for the finite width of the 
analyzed \pt bins~\cite{NIMA.355.541.1995}.  We compare 
the data to inclusive \jpsi data at 200 GeV~\cite{PhysRevLett.98.232002} 
which show similar \pt dependence. At low \pt, the data are compared 
to prompt \jpsi leading-order (LO) NRQCD 
calculations~\cite{PhysRevLett.106.042002,PhysRevD.51.1125} coupled to a 
Color Glass Condensate (CGC) description of the low-$x$ gluons in the 
proton~\cite{PhysRevLett.113.192301}. For the rest of \pt range, the 
data are compared to prompt \jpsi NLO NRQCD 
calculations~\cite{PhysRevLett.106.042002,PhysRevD.51.1125}. The 
LO-NRQCD+CGC calculations overestimate the data at low \pt. The 
NLO-NRQCD calculations underestimate the data at high \pt, while to 
some extent, are consistent with the data at intermediate \pt, 3--5 
GeV/$c$. It is important to stress that the nonprompt \jpsi contribution 
(from excited charmonium states and from $B$-meson decays) 
is not included in these calculations.  This is expected to be a 
significant contribution at high \pt; therefore, the addition of the 
nonprompt \jpsi contribution could account for the difference between 
the data and 
calculations~\cite{PhysRevD.95.092002,PhysRevLett.95.122001,PhysRevC.87.014908}.
The \pt coverage down to zero \pt allows the extraction of the 
\pt-integrated cross section, BR $d\sigma^{J/\psi}_{pp}/dy (1.2<|y|<2.2, 
0<\pt<10~\mbox{GeV/$c$}) =$ 54.3 $\pm$ 0.5 (stat) $\pm$ 5.5 (syst) nb.

 \begin{figure}[ht!]
 \includegraphics[width=1.0\linewidth]{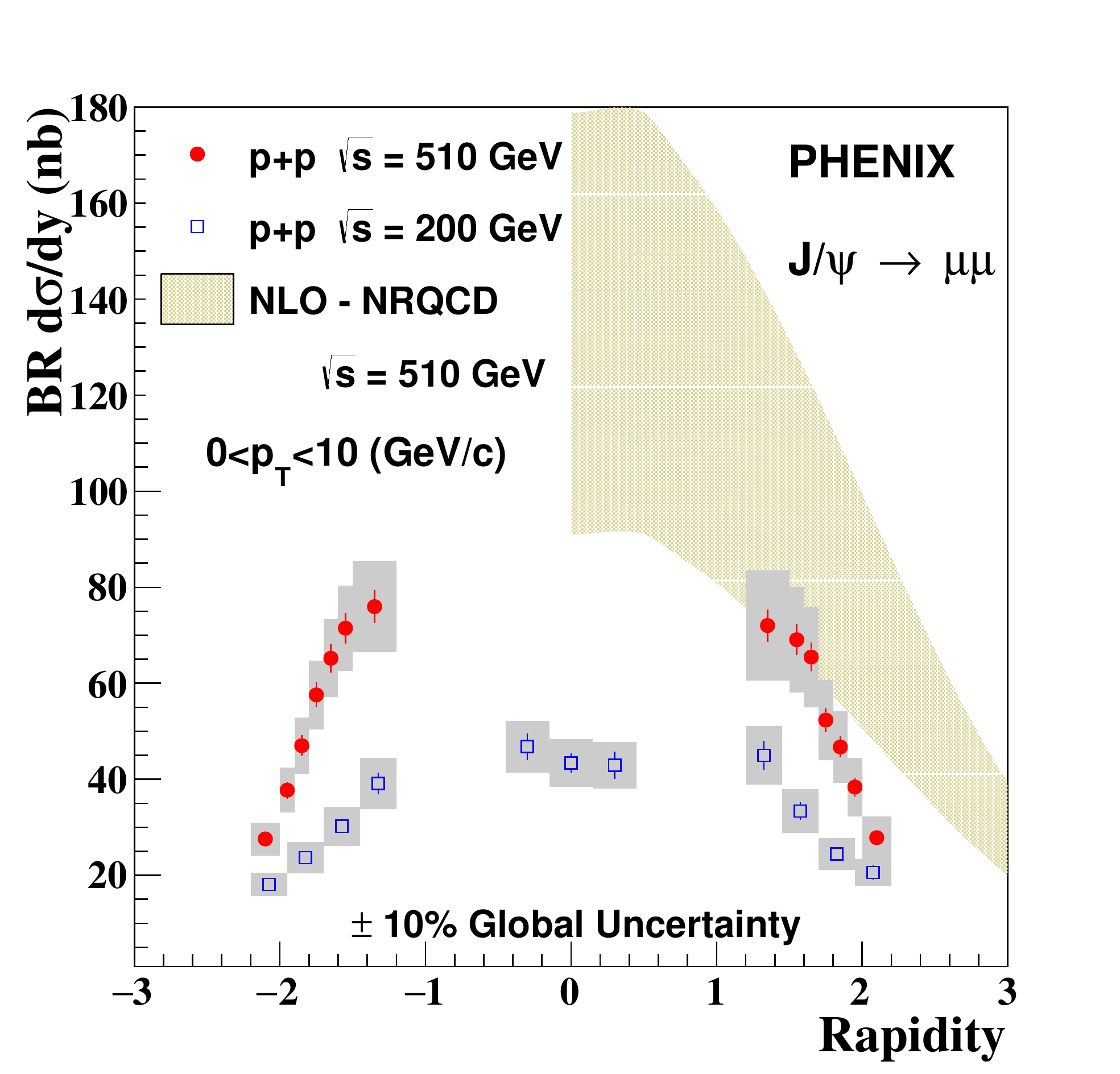}%
\caption{The inclusive \jpsi differential cross section integrated over 
$0<p_T<10$ GeV/$c$ as a function of rapidity at 510 GeV (closed [red] circles) 
and at 200 GeV (open [blue] squares). The error bars 
represent the statistical uncertainties, and the gray shaded band 
represents the quadratic sum of type-B systematic uncertainties. 
NLO-NRQCD calculations~\cite{PhysRevLett.106.042002} are also 
shown.}
\label{fig:yXSjpsi}
 \end{figure}

\begin{table}[htb!]
\caption{Differential cross sections in nb versus rapidity of inclusive 
\jpsi over $0<p_T<10$ (GeV/$c$) with statistical and type-B systematic 
uncertainties.}
\begin{ruledtabular}\begin{tabular}{ccccc} 
& $y^{\rm min}$\ & $y^{\rm max}$ & $BR \frac{d\sigma}{dy}$ (nb) & \\
\hline
& -2.20 \ & \ -2.00 \ & \ $27.6 \pm 1.3 \pm 4.4$ & \\
& -2.00 \ & \ -1.90 \ & \ $37.7 \pm 1.8 \pm 6.0$ & \\
& -1.90 \ & \ -1.80 \ & \ $47.0 \pm 2.1 \pm 7.5$ & \\
& -1.80 \ & \ -1.70 \ & \ $57.6 \pm 2.6 \pm 9.2$ & \\
& -1.70 \ & \ -1.60 \ & \ $65.2 \pm 2.9 \pm 10.4$ & \\
& -1.60 \ & \ -1.50 \ & \ $71.5 \pm 3.2 \pm 11.4$ & \\
& -1.50 \ & \ -1.20 \ & \ $75.9 \pm 3.4 \pm 12.1$ & \\
\\
& 1.20 \ & \ 1.50 \ & \ $72.0 \pm 3.4 \pm 11.5$ & \\
& 1.50 \ & \ 1.60 \ & \ $69.1 \pm 3.2 \pm 11.0$ & \\
& 1.60 \ & \ 1.70 \ & \ $65.5 \pm 3.0 \pm 10.4$ & \\
& 1.70 \ & \ 1.80 \ & \ $52.3 \pm 2.4 \pm 8.3$ & \\
& 1.80 \ & \ 1.90 \ & \ $46.7 \pm 2.2 \pm 7.5$ & \\
& 1.90 \ & \ 2.00 \ & \ $38.4 \pm 1.9 \pm 6.1$ & \\
& 2.00 \ & \ 2.20 \ & \ $27.8 \pm 1.4 \pm 4.4$ & \\
\end{tabular}\end{ruledtabular}
\label{tab:yXSjpsi}
\end{table}

Inclusive \jpsi differential cross section as a function of rapidity 
is listed in Table~\ref{tab:yXSjpsi} and shown in 
Fig.~\ref{fig:yXSjpsi}, which also includes PHENIX inclusive \jpsi data 
at 200 GeV~\cite{PhysRevLett.98.232002} and NLO-NRQCD 
calculations~\cite{PhysRevLett.106.042002}. The 510 GeV data show a 
similar rapidity dependence pattern to that of the 200 GeV data. 
NLO-NRQCD calculations overestimate the data, and this is consistent 
with what was observed in the case of \pt-dependent differential cross 
section (see Fig.~\ref{fig:ptXSjpsi}) because the $y$-dependent 
differential cross section is dominated by the low-\pt region where 
NRQCD calculation overestimates the data. 

\begin{figure}[ht!]
  \includegraphics[width=1.0\linewidth]{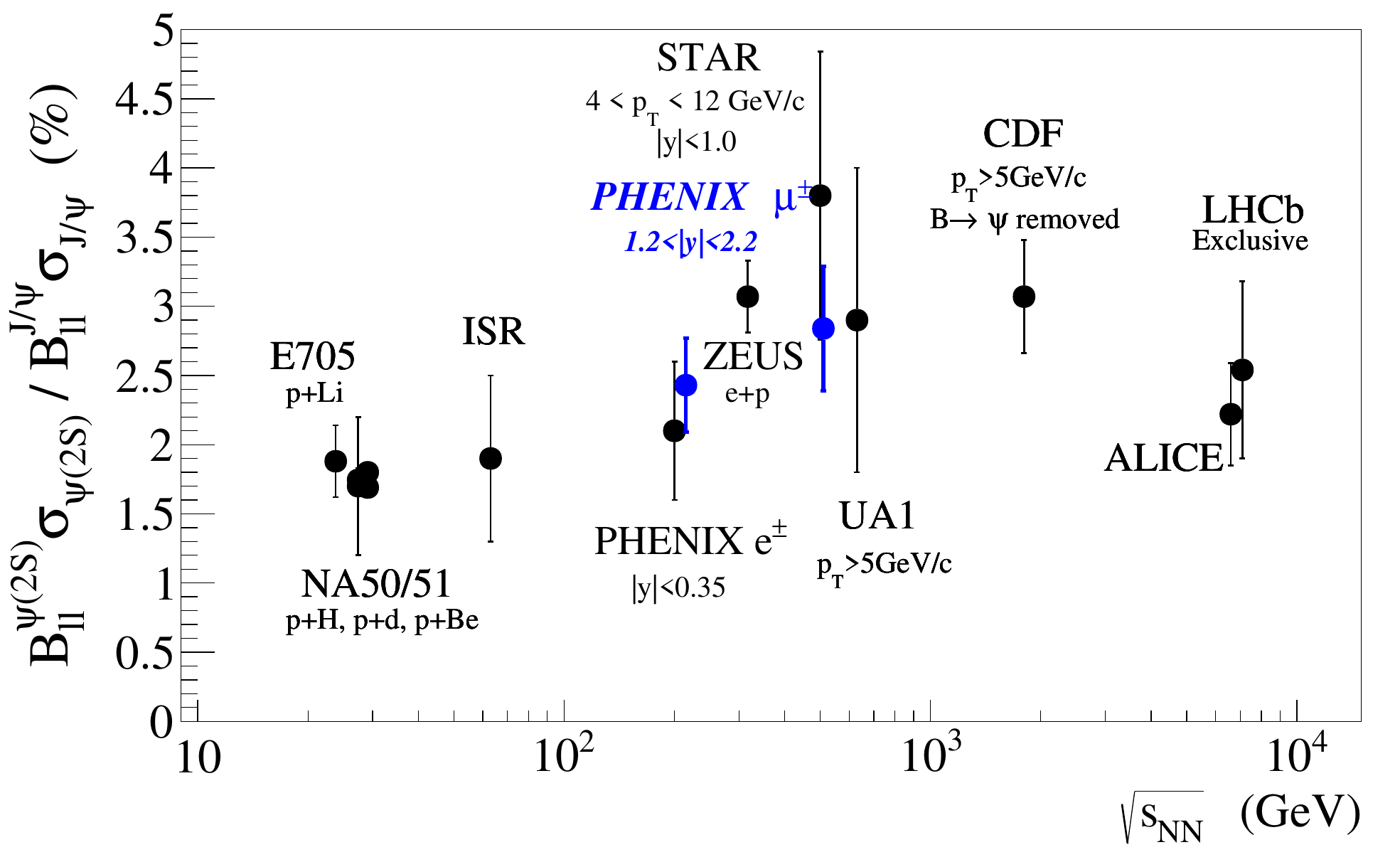}
\caption{Comparison of world data on the ratio of $\psi(2s)$ to \jpsi 
mesons in dilepton 
decays~\cite{EPJC.48.329,PhysRevD.85.092004,PhysRevLett.70.383,PLB.438.35,NPB.142.29,PLB.256.112,PhysRevLett.79.572,JPhysG.40.045001,EPJC.74.2974,PhysRevC.95.034904,PhysRevD.100.052009}. 
The associated uncertainties are the quadrature sum of the statistical 
and systematic uncertainties. 
\label{fig:psiptojpsi}}
 \end{figure}

To quantify the feed-down contribution of excited charmonium states, the 
ratio of the cross section of $\psi(2s)$ to \jpsi, multiplied by their 
respective branching ratio to dimuons, is measured ($R=2.84{\pm}0.45$\%) and 
shown in Fig.~\ref{fig:psiptojpsi}. This ratio is compared with other 
$p$$+$$p$ and $p$$+A$ systems at different collision 
energies~\cite{EPJC.48.329,PhysRevD.85.092004,PhysRevLett.70.383,PLB.438.35,NPB.142.29,PLB.256.112,PhysRevLett.79.572,JPhysG.40.045001,EPJC.74.2974,PhysRevC.95.034904}. 
The results are consistent with world data within uncertainties with no 
significant dependence on collision energy.

To better understand the shape of the \pt spectrum for \jpsi at forward 
rapidity and quantify its hardening at \sqrts = 510 GeV, we calculate 
the corresponding mean transverse momentum $\langle p_T \rangle$ and 
mean transverse momentum squared $\langle p^2_T \rangle$. This is done 
by fitting the inclusive \jpsi \pt-dependent differential cross sections 
with the following 
function~\cite{PhysRevLett.98.232002,EuroPhysJC.77.392}:

 \begin{figure*}[htb!]
\begin{minipage}{0.48\linewidth}
 \includegraphics[width=0.99\linewidth]{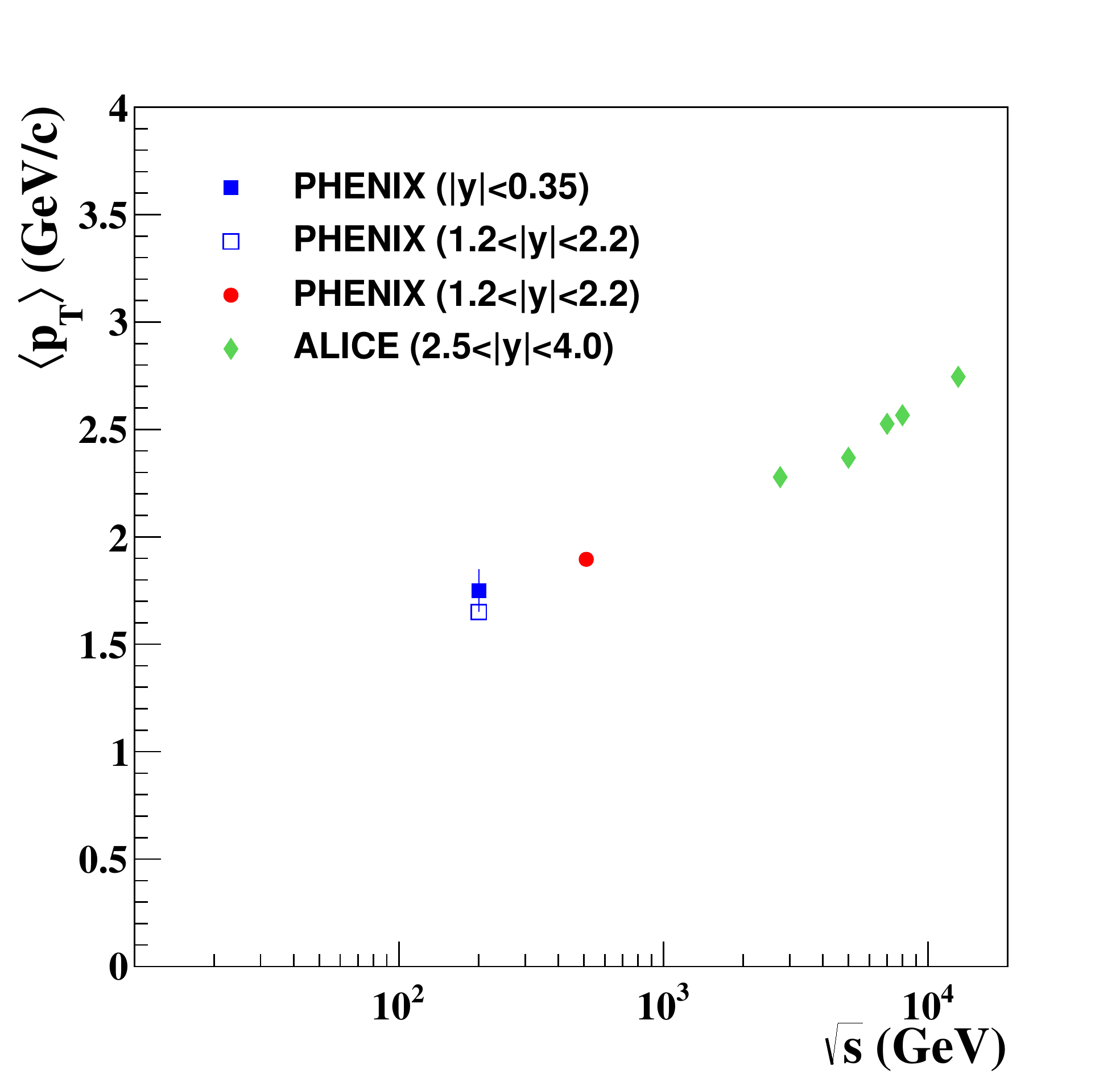}
 \caption{ $\langle p_T \rangle$ as a function of $\sqrt{s}$ for
$J/\psi$ from this work at 510 GeV (closed [red] circle), from PHENIX at 200
GeV~\cite{PhysRevLett.98.232002} at midrapidity (closed [blue] square) and
forward rapidity (open [blue] square), and from ALICE at different
energies~\cite{EuroPhysJC.77.392} (closed [green] diamonds).}
\label{fig:avept1}
\end{minipage}
\hspace{0.2cm}
\begin{minipage}{0.48\linewidth}
 \includegraphics[width=0.99\linewidth]{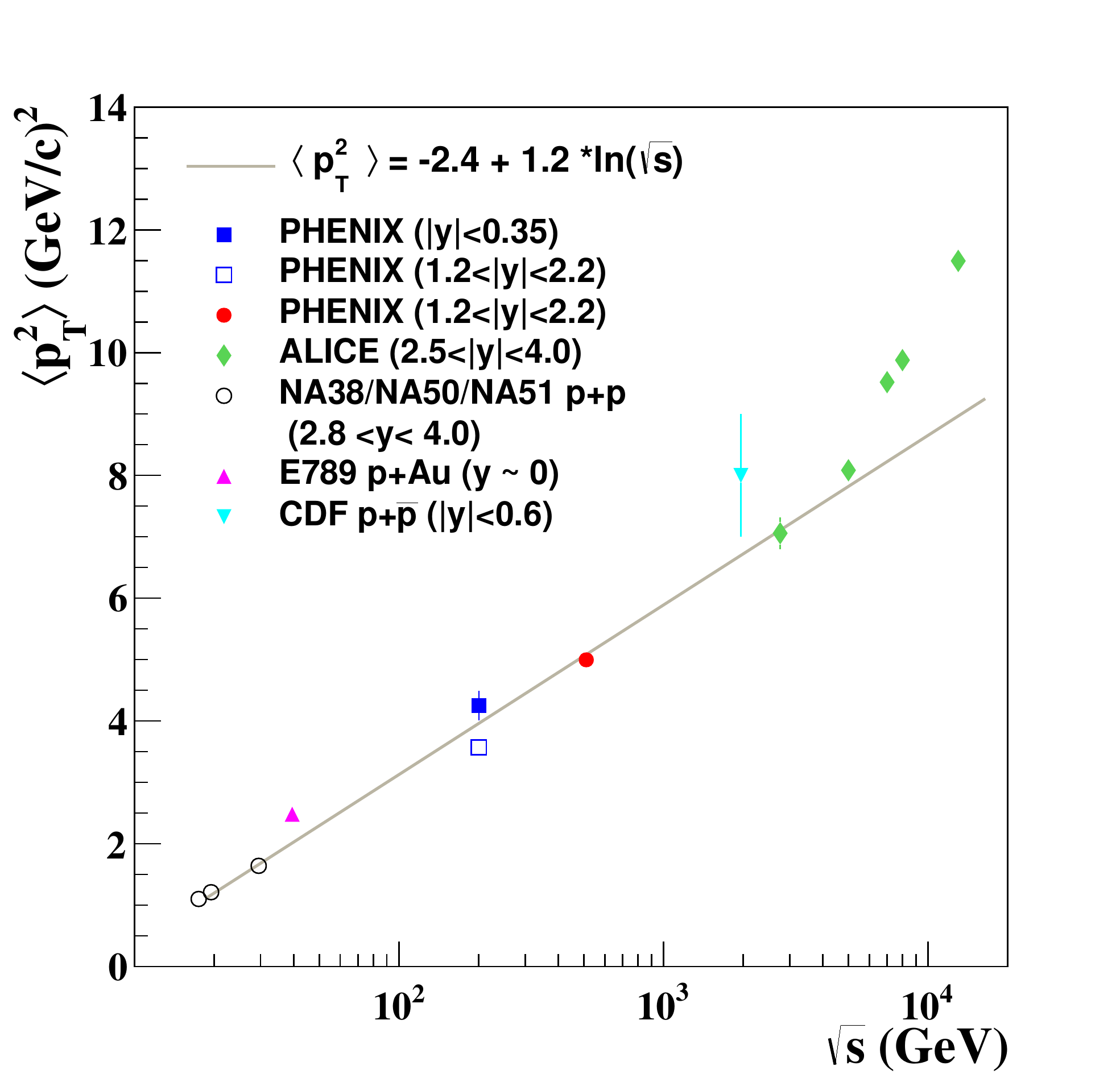}
 \caption{ $\langle p^2_T \rangle$ as a function of $\sqrt{s}$ for
$J/\psi$. The figure includes results from this work at 510 GeV (closed
[red] circle) and PHENIX results at 200 GeV~\cite{PhysRevLett.98.232002} at
midrapidity (closed [blue] square) and forward rapidity (open [blue] 
square).  Also shown are higher-energy data from CDF~\cite{PhysRevD.71.032001} 
(closed [cyan] downward triangle) and ALICE~\cite{EuroPhysJC.77.392}
(closed [green] diamonds). Also shown at lower energies are 
data from NA38/NA50/NA51~\cite{drapier.1998,PhysRevLett.98.232002}
(open [black] circles) and E789~\cite{PhysRevD.52.1307} (closed 
[magenta] upward triangle).
}
\label{fig:avept2}
\end{minipage}
\end{figure*}

\begin{equation}
\label{eq:ptAve}
 f(p_T) = A \frac{p_T}{(1+(\frac{p_T}{B})^2)^n}
\end{equation}
where $A$, $B$ and $n$ are free parameters and their 
values from the fit are $54.6\pm 0.5$, $10.4\pm 0.4$ and $0.45\pm0.06$, 
respectively, and $\langle p_T \rangle$ and $\langle p^2_T \rangle$ are 
the first and second moments of Eq.~\ref{eq:ptAve} in a given \pt range. 
This fit results in a $\langle p_T \rangle = 1.90 \pm 0.02 \pm 0.30$ 
GeV/$c$ and $\langle p^2_T \rangle = 5.00 \pm 0.06 \pm 0.51$ (GeV/$c$)$^2$. 

The first error is statistical, and the second is the systematic 
uncertainty from the maximum shape deviation permitted by the type-B 
correlated errors.

Figure~\ref{fig:avept1} shows $\langle p_T \rangle$ as a function of \sqrts 
from this measurement compared with results from 200 GeV PHENIX data at 
the same rapidity range~\cite{PhysRevLett.98.232002}, and results from 
ALICE at different \sqrts values and in the rapidity range, 
$2.5<y<4.0$~\cite{EurPhysJC.74.2872}. This result follows the increasing 
pattern observed between PHENIX results at 200 GeV and ALICE 
results at 2.76--13 TeV.

Figure~\ref{fig:avept2} shows $\langle p^2_T \rangle$ as a function of 
\sqrts from this measurement compared with several other 
measurements~\cite{PhysRevLett.98.232002,EuroPhysJC.77.392,PhysRevLett.69.3704,PhysRevD.66.092001,PhysRevD.71.032001,EurPhysJC.74.2872,drapier.1998}. 
Similar to $\langle p_T \rangle$, $\langle p^2_T \rangle$ from this 
measurement follows the increasing pattern versus \sqrts established by 
several sets of data over a wide range of energies. Below \sqrts of 2 
TeV, the trend is qualitatively consistent with a linear fit of $\langle 
p^2_T \rangle$ versus the log of the center of mass energy from 
Ref.~\cite{PhysRevLett.98.232002}. However, above \sqrts of 2 TeV, the 
ALICE data indicate $\langle p^2_T \rangle$ grows at an increased rate 
which is interpreted by authors of 
Ref.~\cite{EuroPhysJC.77.392} as due to the fact that ALICE data sets 
have different \pt ranges. The bottom cross section also increases with 
increasing \sqrts, changing the relative prompt and $B$-meson decay 
contributions to the inclusive \jpsi samples discussed 
here~\cite{JHighEnergyPhys.11.065,PhysRevD.95.092002}.  This may also 
contribute to the observed differences in the measured $\langle p^2_T \rangle$.

\begin{figure*}[htb!]
\begin{minipage}{0.48\linewidth}
 \includegraphics[width=0.99\linewidth]{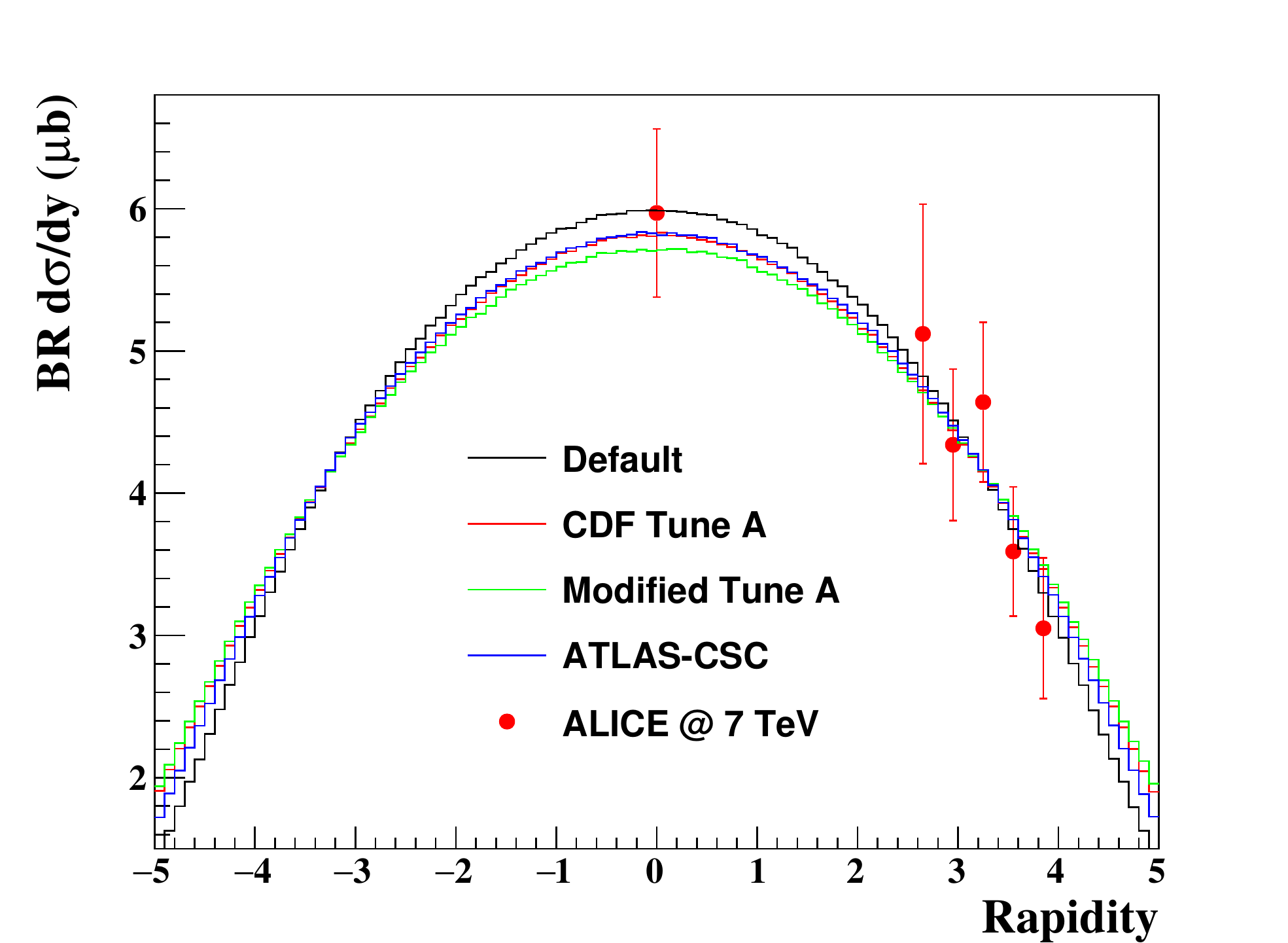}
 \caption{Inclusive \jpsi differential cross section as a function of
rapidity at $\sqrt{s}$ = 7 TeV~\cite{PhysLett.B.704.442} fitted with
different {\sc pythia} tunes at the same
energy.\label{fig:pythiafitat7TeV}}
\end{minipage}
\hspace{0.2cm}
\begin{minipage}{0.48\linewidth}
 \includegraphics[width=0.99\linewidth]{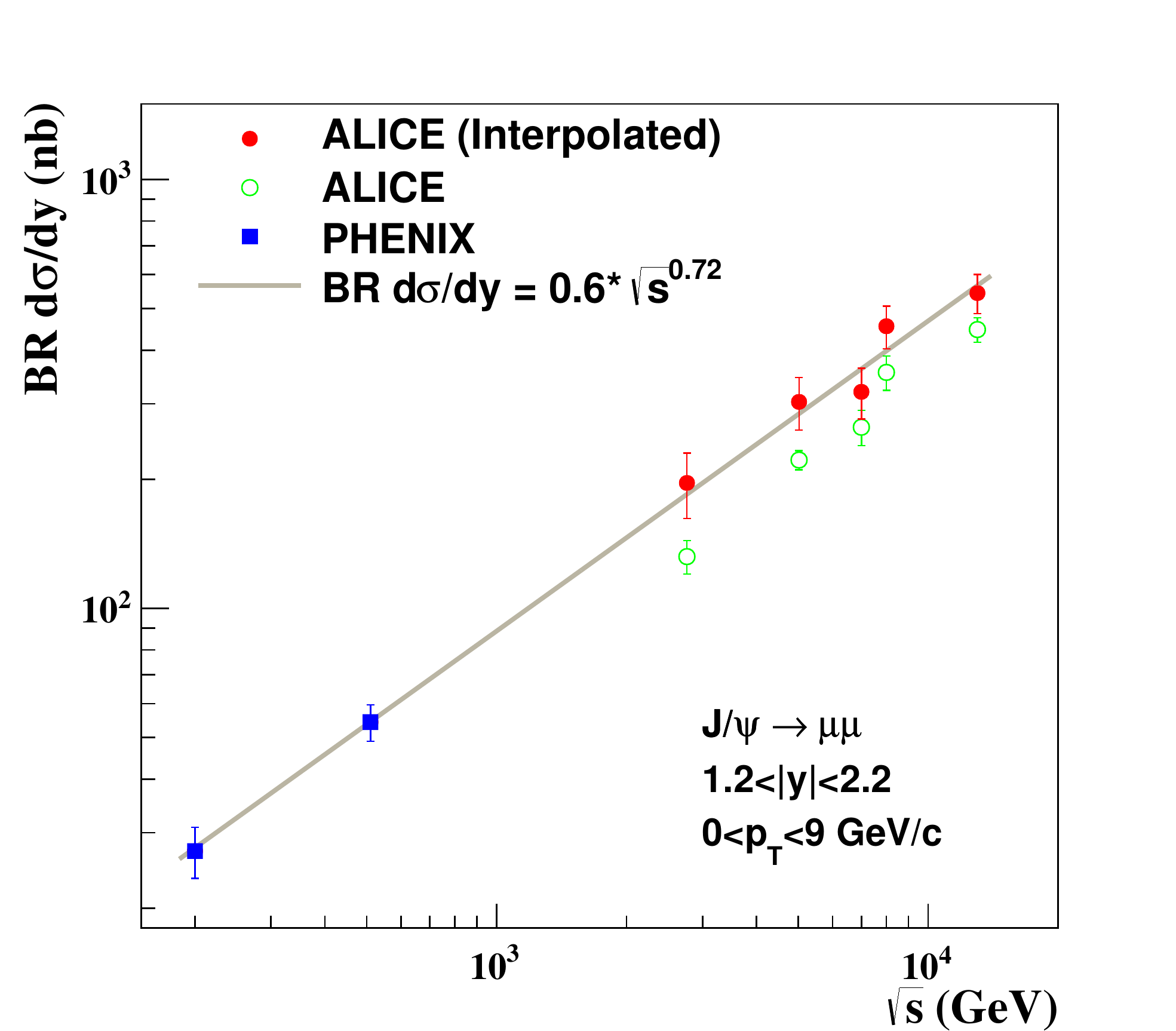}
 \caption{Inclusive \jpsi differential cross section, $d\sigma/dy$, as a
function of \sqrts. The vertical errors correspond to the quadratic sum
of the statistical and type-B systematic
uncertainties.\label{fig:roots}}
\end{minipage}
 \end{figure*}

The $d\sigma^{J/\psi}_{pp}/dy$ measurement at \sqrts= 510 GeV offers an 
opportunity to test the center-of-mass energy dependence of the 
\pt-integrated cross section. Moreover, it bridges the gap between RHIC data 
at 200 GeV and ALICE data starting at 2.76 
TeV~\cite{PhysLett.B.704.442,PhysLett.B.718.295,EurPhysJC.76.184,EuroPhysJC.77.392}. 
However, ALICE data are collected at mid $(|y|<0.9)$ and forward 
$(2.5<y<4.0)$ rapidities and to have a proper comparison we interpolate the 
ALICE data to the PHENIX forward rapidity range, $1.2<y<2.2$. This is done 
by fitting the {\sc pythia} generated $d\sigma/dy$ distribution at each 
energy to the data at the same energy with only the normalization as a free 
parameter. An example is shown in Fig.~\ref{fig:pythiafitat7TeV}. We used 
several {\sc pythia}~\cite{JHEP.2006.026} tunes including PHENIX default, 
tune-A, modified tune-A and {\sc atlas-csc}~\cite{EPJC.39.129}. After 
fitting each of these {\sc pythia} tunes to the data, we extracted 
$d\sigma/dy$ at $1.2<y<2.2$, from these fits. The rms value of the extracted 
$d\sigma/dy$ from the different fits is used in the comparison to RHIC data. 
The error on the rms value is the rms of the errors associated with the fit 
results.

Figure~\ref{fig:roots} shows the results from this measurement, 200 GeV 
PHENIX data (closed [blue] squares), ALICE data (open [green] circles), and 
interpolated ALICE data (closed [red] circles) at several energies. 
Figure~\ref{fig:roots} shows that the data are well described by a power 
law, $d\sigma^{J/\psi}_{pp}/dy \propto (\sqrt{s})^b$, where the exponent 
is $b = 0.72 \pm 0.03$.

\section{Summary}
\label{sec:summ}

We studied inclusive \jpsi production in \pp collisions at \sqrts = 510 
GeV for $1.2 < |y| < 2.2$ and $0 < p_T < 10$ GeV/$c$, through the dimuon 
decay channel. We measured inclusive \jpsi differential cross sections 
as a function of \pt as well as a function of rapidity. The \pt 
integrated differential cross section multiplied by \jpsi branching 
ratio to dimuons is $BR$ $d\sigma^{J/\psi}_{pp}/dy$ $(1.2<|y|<2.2, 
0<\pt<10~\mbox{GeV/$c$}) =$ 54.3 $\pm$ 0.5 (stat) $\pm$ 5.5 (syst) nb. 
With these data measured over a wide \pt range, we calculated $\langle 
p_T \rangle$, $\langle p^2_T \rangle$ and $d\sigma/dy$. The results were 
compared to similar quantities at different energies from RHIC and LHC 
to study their \sqrts dependence. These new measurements could put 
stringent constraints on \jpsi production models.

The inclusive \jpsi differential cross sections were compared to prompt 
\jpsi calculations. These calculations included LO-NRQCD+CGC at low \pt 
and NLO-NRQCD for the rest of \pt range. These model calculations 
overestimated the data at low \pt and underestimated the data at 
high \pt. The nonprompt \jpsi contribution was not included which could 
account for the underestimation at high \pt where the nonprompt 
processes are significant.

In addition, we measured the ratio of the cross section of $\psi(2s)$ to 
\jpsi, multiplied by their respective branching ratio to dimuons, 
$R=2.84\pm0.45$\%. The result is consistent with world data within 
uncertainties with no dependence on collision energy.



\begin{acknowledgments}

We thank the staff of the Collider-Accelerator and Physics
Departments at Brookhaven National Laboratory and the staff of
the other PHENIX participating institutions for their vital
contributions.  We acknowledge support from the
Office of Nuclear Physics in the
Office of Science of the Department of Energy,
the National Science Foundation,
Abilene Christian University Research Council,
Research Foundation of SUNY, and
Dean of the College of Arts and Sciences, Vanderbilt University
(U.S.A),
Ministry of Education, Culture, Sports, Science, and Technology
and the Japan Society for the Promotion of Science (Japan),
Conselho Nacional de Desenvolvimento Cient\'{\i}fico e
Tecnol{\'o}gico and Funda\c c{\~a}o de Amparo {\`a} Pesquisa do
Estado de S{\~a}o Paulo (Brazil),
Natural Science Foundation of China (People's Republic of China),
Croatian Science Foundation and
Ministry of Science and Education (Croatia),
Ministry of Education, Youth and Sports (Czech Republic),
Centre National de la Recherche Scientifique, Commissariat
{\`a} l'{\'E}nergie Atomique, and Institut National de Physique
Nucl{\'e}aire et de Physique des Particules (France),
Bundesministerium f\"ur Bildung und Forschung, Deutscher Akademischer
Austausch Dienst, and Alexander von Humboldt Stiftung (Germany),
J. Bolyai Research Scholarship, EFOP, the New National Excellence
Program ({\'U}NKP), NKFIH, and OTKA (Hungary),
Department of Atomic Energy and Department of Science and Technology
(India),
Israel Science Foundation (Israel),
Basic Science Research and SRC(CENuM) Programs through NRF
funded by the Ministry of Education and the Ministry of
Science and ICT (Korea).
Physics Department, Lahore University of Management Sciences (Pakistan),
Ministry of Education and Science, Russian Academy of Sciences,
Federal Agency of Atomic Energy (Russia),
VR and Wallenberg Foundation (Sweden),
the U.S. Civilian Research and Development Foundation for the
Independent States of the Former Soviet Union,
the Hungarian American Enterprise Scholarship Fund,
the US-Hungarian Fulbright Foundation,
and the US-Israel Binational Science Foundation.

\end{acknowledgments}

%
 
\end{document}